Musical Listening Qualia: A Multivariate Approach

Brendon Mizener[1], Mathilde Vandenberghe-Descamps[2], Hervé Abdi[1], & Sylvie Chollet[2]

[1] University of Texas at Dallas

[2] Junia, Univ. Artois, Université de Liège, Univ. Littoral Côte d'Opale, UMRT 1158 BioEcoAgro, F-62000 Arras, France

Author Note

The authors have no conflict of interest to report. This study was approved by the UT Dallas IRB as initial submission 21-112. Additional materials, including data and stimulus examples, are available at github.com/brendonmiz and at https://osf.io/nkwdc/. The authors would like to thank Pierre Descamps for his help with French translations and musical terminology, and for all of the reviewers for the valuable insight and comments on an initial draft of this paper.

The authors made the following contributions. Brendon Mizener: Stimuli creation, Survey design & creation, Data collection & processing, Statistical analyses, Writing - Original draft preparation, Writing - Review & Editing; Mathilde Vandenberghe-Descamps: Original concept, Survey design & creation, Writing - Review & Editing; Hervé Abdi: Writing - Review & Editing, Statistical guidance; Sylvie Chollet: Original concept, Writing - Review & Editing.

Correspondence concerning this article should be addressed to Brendon Mizener, 800 W. Campbell Rd., Richardson Tex. E-mail: bmizener@utdallas.edu

## List of Figures

## List of Tables

Abstract

French and American participants listened to new music stimuli and evaluated the stimuli using

either adjectives or quantitative musical dimensions. Results were analyzed using Correspondence

Analysis (CA), Hierarchical Cluster Analysis (HCA), Multiple Factor Analysis (MFA), and Partial

Least Squares Correlation (PLSC). French and American listeners differed when they described the musical stimuli using adjectives, but not when using the quantitative dimensions. The present work serves as a case study in research methodology that allows for a balance between relaxing experimental control and maintaining statistical rigor.

*Keywords:* Music Cognition, Multivariate Analyses, Correspondence Analysis, Hierarchical

Cluster Analysis, Multiple Factor Analysis, Partial Least Squares Correlation

Word count: 7271

## Music Listening Qualia: A Multivariate Approach

We have a data collection problem: World Events over the last two years have shown that we need to be able to collect good data outside of the lab. In the lab, because we control error sources, we measure, on relatively small sets of observations, a few well-defined quantitative variables, analyzed using standard techniques such as analysis of variance (ANOVA). But, with the labs closed (remember COVID?), how can we collect good data? Away from the controlled environment of the lab, quantitative variables are hard to measure, but we can collect, on large sets of observations, qualitative variables that can only be analyzed by newer multivariate techniques. In the present paper, we present a case study illustrating this tradeoff.

Something as simple as the sound of a crunch when eating a potato chip can influence its taste (Zampini & Spence, 2004). What about a signal as complex as a string quartet? The present study was designed to quantify a music listening "space" that captures objective stimulus and qualia dimensions to use in future studies investigating cross-modal sensory mapping between food and music.

For the present study, we have defined stimulus dimensions as quantitative musical qualities, such as tempo, range, and meter, and qualia dimensions as qualitative descriptions of music, such as "Dark," "Warm," and "Round." These qualia/qualitative dimensions are similar to the commonly investigated affective or emotional dimensions, but do not specifically assess affective quality. To quantify individual and combined spaces for these concepts, we ran three separate experiments. The first experiment included highly trained musicians and featured a multiple-choice survey about the stimulus dimensions; the second experiment included participants with any level of music training performing a check-all-that-apply task (CATA, Katz & Braly, 1933; Meyners & Castura, 2014; also called "pick any $N$" by Coombs et al., 1956); and the third experiment incorporated both surveys in a single analysis.

To analyze our data, we selected a set of multivariate analyses that can visualize the answers to each of our questions. These multivariate analyses reduce the dimensionality of a data set by computing new variables—called dimensions, components,  factors, or even latent variables—that extract the important information in the table; With these new variables, the original observations (and variables) can be plotted as points in maps that can be interpreted as *conceptual* or *mental spaces* because they represent the similarity among variables or observations by their inter-distances (Abdi & Williams, 2010; Shepard, 1980).

These mental spaces were analyzed with Correspondence Analysis (CA)—a method similar to Principal Components Analysis (PCA)—created to analyze multivariate qualitative data (by contrast with PCA which analyzes quantitative data). We used Multidimensional Scaling (MDS)—a distance analysis method—to visualize the differences between participants and participant groups. To find parallels between the surveys, we used Partial Least Squares Correlation (PLSC)—a method created to analyze two data tables comprising different sets of variables measured on the same set of observations. Finally, we used Multiple Factor Analysis (MFA) to evaluate how French and American participants' responses differed. Each of these methods provided different visualizations and interpretations of the data, which are discussed in more detail below.

### *Music Perception*

Quantifying music perception is an interesting test case for this kind of data gathering and analytical paradigm. Most music or auditory perception studies have the inherent confound that small changes can affect listeners' perception, especially when the study involves timing, tuning, or sound localization. However, the experimental controls may be loosened slightly when investigating holistic music listening, because no single musical element is more important than the whole.

Quantitative and qualitative elements of music are theoretically distinct but practically inseparable (Bruner II, 1990). Listeners respond affectively to technical aspects of music using schemata informed by their individual musical experiences and personality traits (Kopacz, 2005), and composers use various musical and compositional techniques to convey the emotions they want to express (Battcock & Schutz, 2019; Bruner II, 1990). However, quantifying the perceptual interactions between more than one or two musical qualities is a challenge. One reason is that models such as ANOVA and its variations are limited by how many variables a researcher can include while remaining coherent. Another reason is that asking participants to respond to multiple aspects of a stimulus taxes participants' perceptual capacity and is thus difficult to measure (Thompson, 1994).

Music emotion research—in contrast to the research mentioned above—has attempted to capture a more multifaceted perspective on music listening. This is a well-trod domain—see, for example, Juslin and Sloboda (2010)—and the application of multivariate analyses to these questions is similarly well established. Early studies, including Gray and Wheeler (1967) and Wedin (1969, 1972) used MDS to capture the affective space of various musical stimuli. MDS continues to be used commonly in more modern studies (Bigand et al., 2005; Madsen, 1997; Rodà et al., 2014), with a narrow focus on valence and arousal, which were the dimensions originally proposed to be the most salient for perception by Osgood and Suci (1955).

A few studies have specifically investigated dimensions beyond those first two (for example Rodà et al., 2014), and there are conflicting hypotheses about whether the valence-arousal plane or a different model of emotion perception represent the fundamental space around music emotion perception (Cowen et al., 2020; Juslin & Västfjäll, 2008). However, an important distinction between the present study and work in music emotion perception is that the adjectives we chose were informed by music composition and performance, rather than by emotion (Wallmark, 2019).

With regard to musical expertise, many studies evaluate the differences between trained and untrained listeners, but the verdict is still out as to whether trained musicians are better listeners, an issue that could be due to differences about how much training is required for a participant to be "highly trained" (Bigand & Poulin-Charronnat, 2006). There are, however, reported benefits with regard to sensitivity to the emotional content in music (Ladinig & Schellenberg, 2012) and familiarity with tonal systems (Bartlett & Dowling, 1980; Dowling, 1978). Recent works suggest that these benefits may be limited to specific technical aspects, and depend on the extent of training (Raman & Dowling 2017). Although we do not specifically evaluate the differences between trained and untrained listeners in the present study, we included highly trained musicians because they are sensitive to these technical aspects of music and will be able to accurately quantify the stimuli. Additionally, some of the response options to questions on the survey for Experiment 1 would only be familiar to participants with significant music training.

*Intercultural music perception*

There are a few common goals in intercultural studies of music perception. Some quantify the shared emotional experience between musical cultures (Balkwill et al., 2004; Balkwill & Thompson, 1999; Cowen et al., 2020; Darrow et al., 1987; Fritz et al., 2009; Gregory & Varney, 1996), and some ask participants to identify technical aspects of music from other cultures (Raman & Dowling, 2016, 2017). There are fewer studies that include the semantics of language in their evaluation of music perception (Zacharakis et al., 2014, 2015), which makes this topic a prime area for research.

The research program presented in Zacharakis et al. (2014, 2015) deals specifically with timbre perception, and their use of adjectives is similar to how we use adjectives here. In Zacharakis et al. (2014, 2015), Greek and English participants described timbre with adjectives from their native languages. These studies found that while there are some differences, overall, participants' descriptions of timbre do not differ much between languages (Zacharakis et al., 2014, 2015).

### Present questions & methods of analysis

The primary question addressed in this study is: Can we quantify a space around music listening defined by both stimulus and qualia dimensions of music?

Secondary questions include whether French and American participants describe music differently, and whether these differences may arise from cultural differences in music listening or preferences, or are purely semantic. To answer these questions, we employed a set of multivariate analyses that each offered a different perspective on the experimental results. We felt it may be useful to provide a quick overview of the data collection and analytical techniques for readers who may be unfamiliar with these methods.

### Check-all-that-apply (CATA)

The CATA technique—a method widely used in sensory evaluation (and elsewhere but under different names: Coombs et al., 1956; Katz & Braly, 1933; Meyners & Castura, 2014; Valentin et al., 2012)—measures how participants describe a set of stimuli. In a CATA task, stimuli are presented one at a time, and for each stimulus, participants are shown a list of descriptors and are asked to select the descriptors that apply to the presented stimulus (Meyners & Castura, 2014). CATA easily assesses questions with multiple "correct" responses (Coombs et al., 1956), and places little cognitive demand on participants because they do not have to generate responses (Ares et al., 2010).

CATA data are analyzed by 1) computing a pseudo contingency table that records the number of times descriptors were associated with stimuli and 2) analyzing this data table with Correspondence Analysis in order to visualize the patters of association between a) stimuli, b) descriptors, and c) stimuli and descriptors.

*Correspondence Analysis*

The primary analysis used on the stimulus response data collected in the surveys is Correspondence Analysis (CA, Benzécri, 1973; Escofier-Cordier, 1965; Greenacre, 1984). CA is similar to Principal Components Analysis (PCA) but is performed on qualitative data. Specifically, just like PCA, CA analyzes a contingency table by computing components (whose number is the lesser of $I-1$ and $J-1$, where $I$ is the number of rows and $J$ is the number of columns) that capture the relationships within, respectively, the rows (observations) and columns (variables) of the data table—in our case between musical excerpts and descriptors. In CA, the components for the rows and the columns have the same variance and can therefore be visualized in the same space. This makes CA a method of choice when the experimental questions investigate how all variables *and* observations are related to one another.

*Hierarchical Cluster Analysis*

Hierarchical Cluster Analysis (HCA, Pielou, 1984) identifies groups, or clusters, of observations from the rows of a distance matrix, because HCA displays these observations as "leaves" on a tree computed to best represent the original distances. This method was used here to determine whether there were clusters of excerpts or adjectives that arose during participant ratings. These clusters were used as design or grouping variables and to select colors for visualizations.

*Multidimensional Scaling*

Metric Multidimensional Scaling (MDS, Abdi, 2007; Borg & Groenen, 2005; Gower, 1966; Hout et al. 2013; Kruskal & Wish, 1978; Torgerson, 1958; Shepard, 1962)—a technique commonly used in music perception studies (Bigand et al., 2005; Rodà et al., 2014; Wedin, 1969, 1972)—analyzes a distance matrix computed between observations and visualizes these observations by positioning them on a map such that the distance between observations on the map best approximates their distance in the data table. MDS is commonly used to represent the similarity between stimuli; here, this technique is used to evaluate the similarity between groups of participants.

*Multiple Factor Analysis*

Multiple Factor Analysis (MFA, Abdi et al., 2013; Escofier & Pagès, 1994) extends PCA to analyze and visualize multiple tables or blocks of variables that each describes the same observations. MFA computes a *compromise* and a set of *partial factor scores*, where the *compromise* is the average of (or compromise between) the normalized factor scores from each block, and the *partial factor scores* are the factor scores of each individual block. Plotting these factor scores allows for the comparison of observations (rows), and, for each observation, the relationships between the blocks of variables that contributed to that observation. The basic difference between MFA and PLSC is that PLSC extracts commonalities between two data tables, whereas MFA extracts similarities and differences between two or more data tables.

In the present study, MFA was used to evaluate differences between French and American participants in how they described specific excerpts and used specific adjectives. This application of MFA can be generalized to different groups of participants or other sources of data measured on the same set of observations. When the data take the form of a contingency table, MFA allows for the analysis of the contributions to both the observations and the variables.

*Partial Least Squares Correlation*

Partial Least Squares Correlation (PLSC, Abdi & Williams, 2013; Tucker, 1958) analyzes two data tables that describe the same set of observations (rows) with two different sets of variables (columns). To extract the common information between the two data tables, PLSC separately combines the variables from each data table to create new variables—similar to factor scores and called *latent variables*—that have the largest covariance. This method is commonly used in neuroimaging studies to extract the common information between imaging and behavioral data (Krishnan et al., 2011). It was used in the present study to evaluate the similarities in how participants in either survey rated the excerpts.

*Bootstrapping*

We use bootstrapping (Hesterberg, 2011) to evaluate group differences because the methods outlined above are not inferential methods, and do not inherently allow for hypothesis testing. Bootstrapping evaluates the stability of the result of an experiment. This is displayed in the form of confidence intervals, as ellipses, in the plots below, computed from resampling the original observations (Hesterberg, 2011).

*Permutation testing*

We used permutation testing (Berry et al., 2021) to evaluate the significance of results of the analyses described above. Permutation testing compares the signal present in the observed data to permutations of these data and computes test statistics on each permutation. The test statistic of the observed data is then evaluated relative to the distribution of test statistics from the permuted data. The extremity of the observed values—e.g., the most extreme 5% for a $p < .05$ significance level—indicates the significance of the signal in the data.

<div align="center">

**Experiment 1:  Musical Qualities Survey**

</div>

## Methods

### *Participants*

For the first experiment, we recruited highly trained musicians with a minimum of 10 years of formal music training to evaluate the stimulus dimensions or musical qualities, and to ascertain whether these stimuli truly reflected the composer's intent of varying on a wide range of musical dimensions (Raman & Dowling, 2017). Participants in the United States and in France were recruited by word of mouth and social media. There was a total of 84 responses to the survey, of which 57 were removed for not completing the survey, leaving a total of 27 ($N_{\text{France}} = 9$, $N_{\text{USA}} = 18$) for the analysis. All recruitment measures were approved by the UT Dallas IRB.

### *Stimuli*

All stimuli were new, original excerpts composed—in various Western styles using Finale composition software (Finale v25, MakeMusic, Inc.)—by the first author specifically for this study (scores and audio files available upon request). Each stimulus was a wav file generated using the Finale human playback engine, approximately 30 s in length (range: 27 – 40 s, $M = 32.4$ s). The stimuli were all string quartets, a choice made to control for effects of timbre but also vary a number of musical qualities, specifically: Harmony, Tempo, Meter, Density, and Genre.

### *Survey*

American and French participants received links to surveys presented via Qualtrics in (respectively) English and French. Participants were instructed to listen to the excerpts presented in the survey using headphones or in a quiet listening environment, but this was not controlled, nor was it assessed as part of the survey. After standard informed consent procedures, participants

Table 1

*Musical Qualities and the provided survey response options.*

| Harmonic Material | Tempo | Meter | Density | Genre | Dynamics |
|---|---|---|---|---|---|
| Diatonic: Major | Very slow | Simple Duple | Very sparse | Baroque | Soft |
| Diatonic: Minor | Slow | Simple Triple | Moderately sparse | Classical | Moderate |
| Blues | Moderately Slow | Simple Quadruple | More sparse than dense | Romantic | Loud |
| Chromatic | Moderate | Compound Duple | More dense than sparse | Impressionist | Varied: gradual crescendo |
| Whole tone | Moderately Fast | Compound Triple | Moderately Dense | Modern | Varied: gradual decrescendo |
| Modal | Fast | Compound Quadruple | Very Dense | Jazz/Blues | Some of each, soft and loud |
| Quintal/Quartal | Very Fast | Complex | | Contemporary | |
| Ambiguous | | | | Other | |
| Other | | | | | |

| | Contour | Motion | Range | Articulation |
|---|---|---|---|---|
| | Ascending | Conjunct | Narrow | Staccato |
| | Descending | Disjunct | Moderate | Marcato |
| | Arch | Combination of conjunct | Wide | Legato |
| | Undulating | and disjunct | Very Wide | Tenuto |
| | Pendulum | I do not think this | I do not think this | Other |
| | Terrace | excerpt has a melody | excerpt has a melody | |
| | I do not think this | Other | | |
| | excerpt has a melody | | | |
| | Other | | | |

listened to 15 of the 30 excerpts, presented one at time in a random order, and answered ten questions per excerpt, one for each of the musical qualities being assessed. The musical qualities assessed and the levels associated with each quality are shown in Table 1. Of these dimensions being assessed, all were multiple choice, allowing for a single response, except for meter, contour, and articulation, which were check-all-that-apply (See supplementary materials for the French translations of this table). Upon completion of the experimental task, participants were asked to provide demographic data, including age, gender identity, nationality, occupation, and musical experience.

### Data Processing

To process the data, survey responses were converted into a "brick" of data, with the excerpts on the rows, the qualities on the columns, and the participants on the pages (See Figure 1). For the current experiment, one quality is one variable, and we refer to the response options as levels of that variable. On any page, at the intersection of any row and column was a one or a zero, with a one indicating that this participant had selected this level of this quality (column) to describe this excerpt (row). The responses in the French "brick" were all translated into English, and then the bricks from both nationalities were summed together across pages to obtain a single pseudo-contingency table[1] in which the intersection of a row and a column was the total number participants who selected a level of a musical quality to describe an excerpt. Levels for which the column sum was equal to one were considered as outliers and removed from the data.

After removing these columns, preliminary visualizations revealed a few variables that required recoding because they were having an outsized effect on the analysis. For the "Meter," variable, there were initially seven levels: "Simple Duple," "Simple Triple," "Simple Quadruple," "Compound Duple," "Compound Triple," "Compound Quadruple," and "Complex," but some participants misunderstood this question and selected multiple options for each level of "Duple," "Triple" or "Quadruple." These responses were recoded, removing "Simple," "Compound" and "Complex," (there were no excerpts with complex meter), and collapsed, so that each excerpt had only one meter response per participant, "Duple," "Triple," or "Quadruple."

---

[1] In a real contingency table the observations are independent of each other and therefore one observation contributes to one and only cell of the table. By contrast with CATA one respondent provides a *set* of responses that therefore contributes to *several* cells of the data table—a pattern that breaks the independence assumption.

*Figure 1*. Survey data processing flowchart. In the top table, participants are in rows and excerpts are in blocks of columns. Purple cells indicate that participants were presented with and responded to an excerpt, gray cells indicate that participants were not presented with an excerpt.

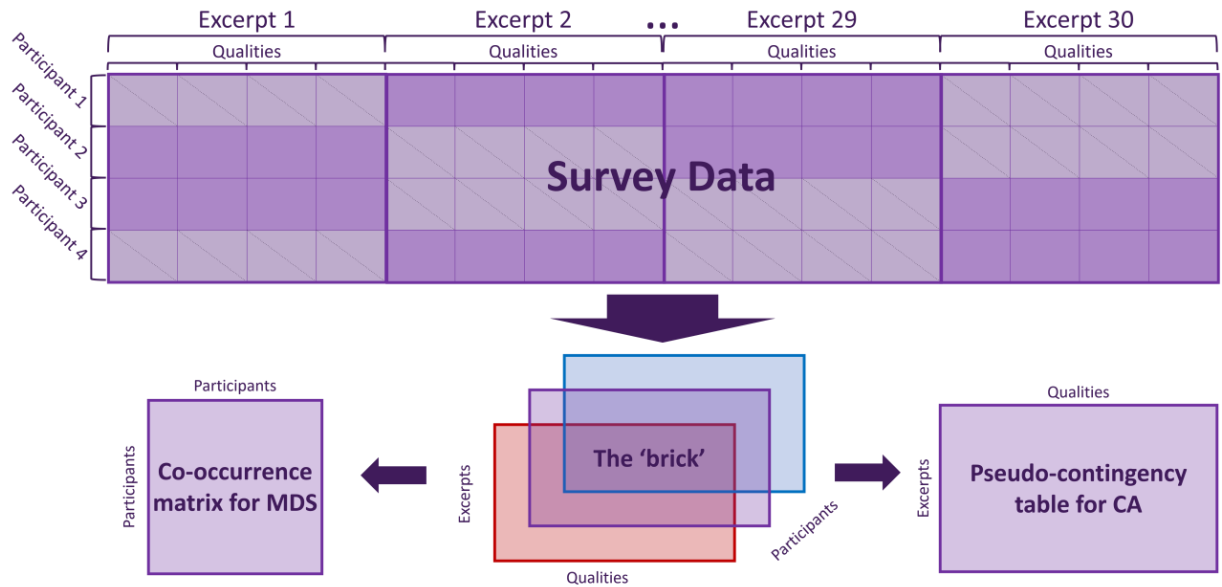

There were multiple qualities for which a possible response was "I do not think this excerpt has a melody," and this pattern created a problem in which multiple columns represented the same response, which had a similar effect to the one caused by the "Meter" variable before it was recoded. To avoid this problem, these responses were also recoded. A new variable, "Melody," was created, with two levels/columns, yes and no, and if participants responded "I do not think this excerpt has a melody" to any of the Contour, Motion, or Range variables, a one was counted in the "No" column for that participant and that excerpt. The other levels for each of these three variables were then recoded so that each other column for that variable in that row had the value of one divided by the number of options for that variable—a procedure called barycentric recoding. If the participant responded with "I do not think this excerpt has a melody" for some but not all three of those variables, a one was still counted in the "no melody" column, but only the variables for which "I do not think…" was selected were recoded using barycentric recoding. For all excerpts and participants for which "I do not think…" was never selected, a one was added to the "Yes" column for the melody variable. Once the data were recoded, the brick was once again summed across pages to obtain the data table that would be used for subsequent analyses.

### Analysis

To analyze the similarity structure between participants, we computed a co-occurrence matrix from the brick with participants on the rows and columns, such that the intersection of a row and column represented the number of common choices between participants. This co-occurrence matrix was then analyzed using MDS.

To analyze the excerpts and musical qualities and obtain the music quality space, we performed a CA on the contingency table. To identify potential clusters among the excerpts, we ran an HCA on the row factor scores obtained from the CA.

## Results

### Participants

The MDS performed on the co-occurrence matrix of participants was intended to identify potential clusters of participants. Visual examination of the results did not reveal any clusters—a pattern suggesting that the participants constituted a homogeneous group. To confirm this conclusion, we also computed average factor scores by nationality and gender identity and bootstrap-derived confidence intervals around these averages and did not find any significant differences (See supplementary materials for plots).

*Excerpts*

The results of the CA and subsequent permutation testing performed on the contingency table revealed 18 significant dimensions. In such a scenario, it is important to remember that significant is not synonymous of important, and the dimensions we consider are limited by interpretability. For the current study, we have focused on the first two dimensions, which account together for 32.74% of the total variance. Figure 2 displays the scree plot, which shows for this analysis the percentage of variance explained by each dimension. Readers curious about dimensions three through five are recommended to the supplementary materials.

*Figure 2.* CA: Scree plot for the Qualities Survey, showing percentage of explained variance per dimension. The horizontal line indicates the average variance extracted per dimension. Purple dots indicate significant dimensions as determined by permutation testing.

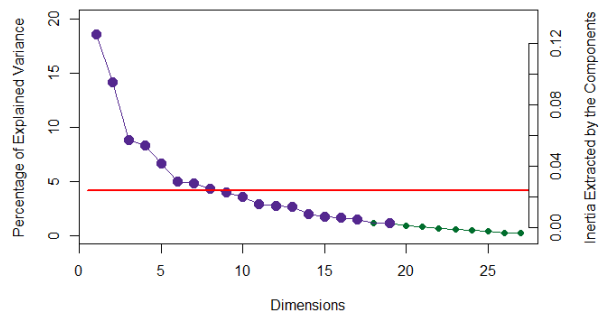

Preliminary plots of the factor scores obtained from the CA revealed that Excerpts 6 and 14 distorted the factor space, with these two excerpts dominating the second and third dimensions. To help interpret the factor space, these two excerpts were removed from the data and the CA was rerun. Excerpts 6 and 14 were then added back in as *supplementary observations* (see Abdi & Béra, 2018, for details), a technique which visualizes the information that these elements share with the elements retained in the main sample without distorting the factor space. The proximity of these supplementary observations to the origin helps identify how much information is shared with the rest of the sample. The closer the supplementary observations are to the origin, the less information they share with the rest of the sample.

The HCA performed on the row factor scores of the CA revealed four clusters of excerpts (see supplementary materials for the tree diagram). Figure 3 displays the first two dimensions for the row factor scores of the CA, colored according to the clusters revealed by the HCA, with Excerpts 6 and 14 as supplementary observations colored separately. Figure 3 also displays the column factor scores for the qualities calculated by the CA (right), with the levels of a given quality colored the same. For clarity, only the levels of qualities that contributed more than the average (see below, and Figure 4) are displayed.

*Figure 3.* CA: Musical Qualities Survey, factor plots for Excerpts, colored according to clusters identified by the HCA, and important musical qualities, colored such that levels of each quality are the same color. Axes are labeled with the dimension, eigenvalue, and the explained variance for the dimension.

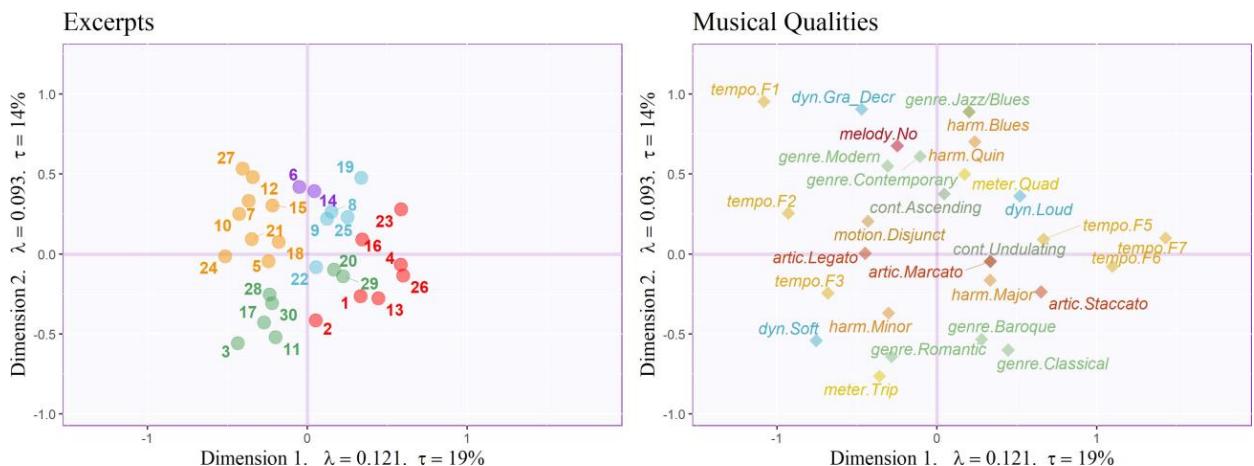

The proximity between two points in Figure 3 indicates their similarity when these points are on the same map. Because the CA computes a space common to both rows and columns, points on different maps can also be compared. Proximity between points on separate maps reflects their association relative to the average, for example Excerpt 24 is more associated with Legato articulation than is the average excerpt (Abdi & Williams, 2010).

        To evaluate the relative importance of the excerpts and musical qualities in defining each dimension, we computed their *contributions* to the dimensions. Contributions are similar to squared coefficients of correlation and vary between zero and one with zero indicating no importance and one indicating maximum importance (Abdi & Williams, 2010). Contributions—being squared—are positive, but to facilitate interpretation contributions are signed to express the sign of the corresponding factor scores. Contributions whose magnitude is larger than the average contribution (i.e., 1 divided by the number of scores) are considered important for their factorial dimensions. A plot of the contributions for all excerpts and variables is in the supplementary materials.

*Figure 4.* CA: Musical Qualities survey, important signed contributions for the first two dimensions, colored similarly to Figure 3. The *y*-axis represents the value of the contributions.

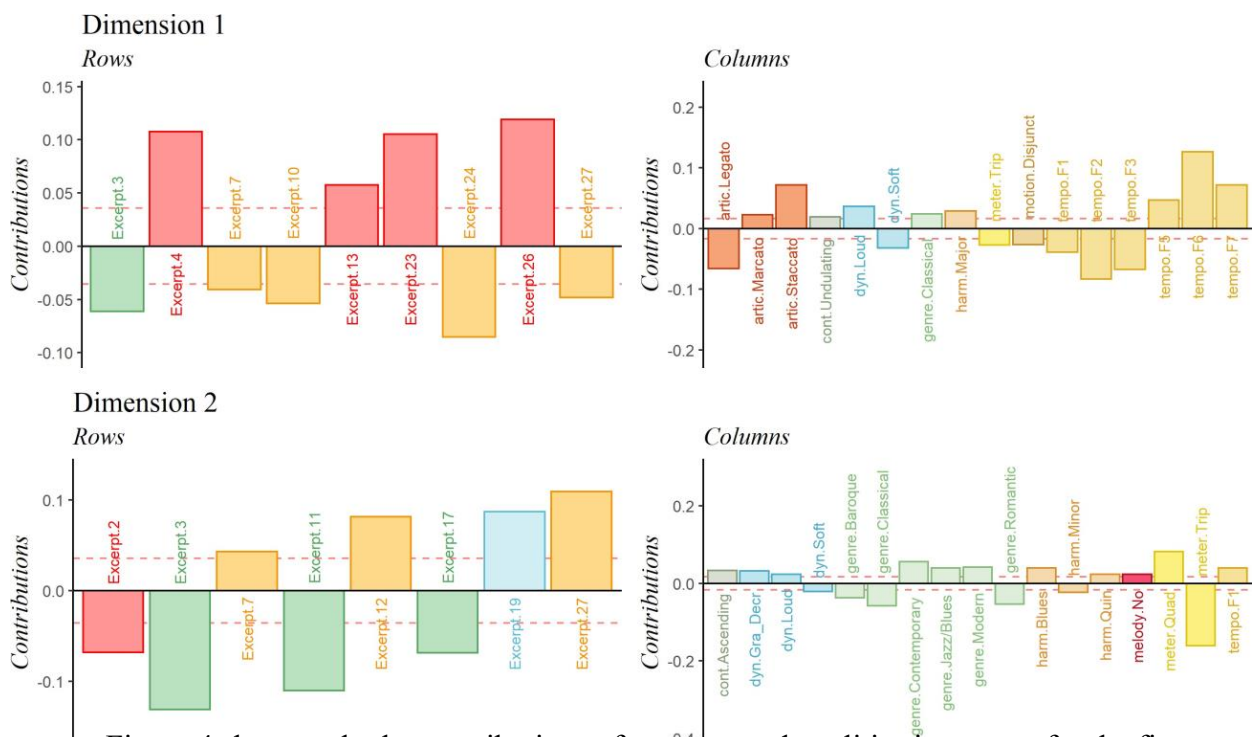

Figure 4 shows only the contributions of excerpts and qualities important for the first two dimensions. Tempo, articulation, and dynamics contribute importantly to the first dimension, along with  a few single levels from other variables. Genre and meter, and to a lesser extent harmony, dynamics, and contour all contribute to the second dimension. For both dimensions, the excerpts that are associated with these levels of variables also contribute importantly.

**Experiment 1 Discussion**

Observing no group differences between French and American participants in the results of the MDS on the co-occurrence matrix suggests that the trained musicians perceived the excerpts with which they were presented similarly.

The results of the CA (Figure 3) reveal a few musical connections: for example, between tempo and articulation (on Dimension 1), and between genre and harmony (on Dimension 2). Staccato articulations, associated on this factor plot with high tempi, are played light and separate, and legato articulations, associated with slow tempi, are played smooth and connected. The coordinate mapping of jazz/blues harmony and genre, which are on top of one another, is the most extreme example of a genre being associated with certain harmonic material, but other connections are also revealed. The second dimension separates older styles, such as Baroque, Classical, and Romantic, from newer styles, Contemporary, Jazz/Blues, and Modern. Dimension 2 similarly separates harmonies associated with those styles, specifically older and simpler harmonies of major and minor from the more complex harmonies of Quintal and Blues.

The first dimension can be interpreted as arousal—tempo, articulation, and dynamics all load from greater arousal (i.e., higher tempos, greater dynamics) in the positive direction to lesser arousal (i.e., lower tempos, softer dynamics) in the negative direction on the first dimension. Dimension 2 is less clear and does not seem to be tied to valence. Minor and major harmony, for example, both score negatively on Dimension 2. Instead, Figure 4 shows that while two levels of the meter variable are the most important for this dimension, that genre is also important, based on the number of levels of genre that contribute significantly to Dimension 2. Considering the contributions of the genre and the harmony variables, it may be that the second dimension represents complexity.

## Experiment 2: Musical Adjectives Survey

## Methods

### *Participants*

Participants with self-reported normal hearing were recruited for Experiment 2 without regard to level of music training. Participants in the United States were recruited by the UT Dallas Psych Research Sign-up (SONA) System, by word of mouth, and by social media. French participants were recruited by word of mouth, email, and social media. Only participants who signed up via the SONA System were compensated (i.e., with research participation credit). Other participants—including all French participants and US participants who did not sign up via the SONA System—were not compensated. Out of 520 survey responses received, 167 were incomplete and removed. The remaining 354 were filtered by nationality: American participants who answered the question "What's your nationality?" with a compound nationality including American were retained, but those who indicated only a nationality other than American were excluded. For example, "Indian-American" was included, but "Ghanian" was not. This left a total of 277 ($N_{France} = 111$, $N_{USA} = 166$) survey responses for analysis. Table 2 shows further demographics including all self-identified nationalities. All recruitment measures were approved by the UT Dallas IRB.

### *Stimuli*

The stimuli used for Experiment 2 were the same as those used for Experiment 1.

Table 2

*Participant Demographic Data*

| | | Experiment 1 | | |
|---|---|---|---|---|
| Nationality | Gender identity | Age (years) | Age range | Years of Training |
| France | F ($N = 4$) | $M = 41.25$, $SD = 13.59$ | $28 - 60$ | $M = 15.50$, $SD = 3.32$ |
| | M ($N = 5$) | $M = 32.0$, $SD = 2.73$ | $29 - 36$ | $M = 16.80$, $SD = 6.30$ |
| US | F ($N = 7$) | $M = 27.71$, $SD = 10.7$ | $19 - 49$ | $M = 17.14$, $SD = 12.38$ |
| | M ($N = 11$) | $M = 30.91$, $SD = 11.69$ | $19 - 61$ | $M = 18.64$, $SD = 11.58$ |

All reported nationalities:

| | |
|---|---|
| France | French |
| US | American |

| | | Experiment 2 | | |
|---|---|---|---|---|
| Nationality | Gender identity | Age (years) | Age range | Years of Training |
| France | F ($N = 72$) | $M = 20.83$, $SD = 4.36$ | $18 - 52$ | $M = 3.40$, $SD = 4.01$ |
| | M ($N = 35$) | $M = 20.14$, $SD = 1.77$ | $18 - 24$ | $M = 4.60$, $SD = 4.88$ |
| | Non-Binary/Did not disclose ($N = 4$) | $M = 20.25$, $SD = 0.96$ | $19 - 21$ | $M = 3.25$, $SD = 2.62$ |
| US | F ($N = 102$) | $M = 22.11$, $SD = 5.31$ | $18 - 51$ | $M = 3.32$, $SD = 3.41$ |
| | M ($N = 61$) | $M = 22.32$, $SD = 5.21$ | $18 - 54$ | $M = 2.98$, $SD = 3.55$ |
| | Non-Binary/Did not disclose ($N = 3$) | $M = 19.67$, $SD = 1.53$ | $18 - 21$ | $M = 5.33$, $SD = 4.16$ |

All reported nationalities:

| | |
|---|---|
| France | French, French-Belgian |
| US | American, Asian American, African American, Brazilian-American, Bengali-American, Chinese, Egyptian-American, El Salvadoran, Ethiopian-American, Indian-American, Indian, Italian, Kurdish, Mexican, Mexican-American, Moroccan, Nigerian-American, Pakistani, Pakistani-American, South Korean, Sri Lankan, Turkish, Vietnamese |

*Survey*

The procedures for participants in Experiments 1 and 2 were similar: American and French participants received links to surveys presented via Qualtrics in (respectively) English and French., and instructions regarding listening environment were the same as in Experiment 1. After standard informed consent procedures, participants listened to 15 of the 30 excerpts, presented one at a time, in a random order, and performed a CATA task. Participants were instructed to select all adjectives that they felt described the stimulus. Participants were provided with a list of 33 adjectives, presented in a random order for each stimulus, such as such as "Dark," "Warm," and "Colorful" (French: "Sombre," "Chaleureux," and "Coloré"). The adjectives for this survey were selected using Wallmark (2019) as a guide and in consultation with a French professional musician. Some adjectives were initially selected in French and some in English. In all cases, adjectives were selected for which there was a clear French (vis-à-vis English) translation. The adjectives are listed in English and in French in the supplementary materials. Following the experimental task, the participants were asked to provide demographic data, including age, gender identity, nationality, occupation, and musical experience.

*Data Processing & Analysis*

Data for the survey for Experiment 2 were processed similarly to Experiment 1. Due to a technical error, French participants were not presented with Excerpt 17, so the data for that excerpt were removed from the dataset for the American participants. Although Excerpts 6 and 14 were removed from Experiment 1 data for being outliers, they were not found to be outliers in Experiment 2, and were, therefore, included in all analyses for this experiment. To process the data, first, all French survey responses were translated into English. Both sets of responses were then converted into "bricks" of data, with the excerpts on the rows, the adjectives on the columns, and participants on the pages. On a page, at the intersection of a row and column was a one or a zero, with a one indicating that this participant had selected this adjective (column) to describe this stimulus (row). The bricks were then summed across pages to obtain a pseudo-contingency table in which the intersection of a row and a column stored the number of participants who selected an adjective to describe an excerpt.

To analyze the similarity structure between participants, we computed a co-occurrence matrix from the brick with participants on the rows and columns, such that the intersection of a row and column represented the number of common choices between participants. This co-occurrence matrix was then analyzed using MDS.

To analyze the excerpts and adjectives and obtain the music quality space, we performed a CA on the excerpts by adjectives contingency table. To identify potential clusters of excerpts or adjectives, two separate HCAs were computed, one on the row factor scores (excerpts) and one on the column factor scores (adjectives) obtained from the CA.

We performed two MFAs, one to explore differences between French and American participants from the perspective of their use of adjectives, and another to explore differences between French and American participants from the perspective of their descriptions of the excerpts. To prepare the data for these MFAs, we separated the brick into two separate bricks, one for the French participants and one for the American participants. Each brick was then summed to obtain excerpts by adjectives pseudo-contingency tables for each nationality. These tables were then transposed to obtain adjectives by excerpts pseudo contingency tables for each group. The French and American excerpts by adjectives tables were then concatenated into a single large matrix in which each table represented a block, as were the transposed (adjectives by excerpts) tables. We then performed separate MFAs on each of these large matrices. Figure 5 sketches the data manipulation process.

*Figure 5.* MFA: Data manipulation flowchart.

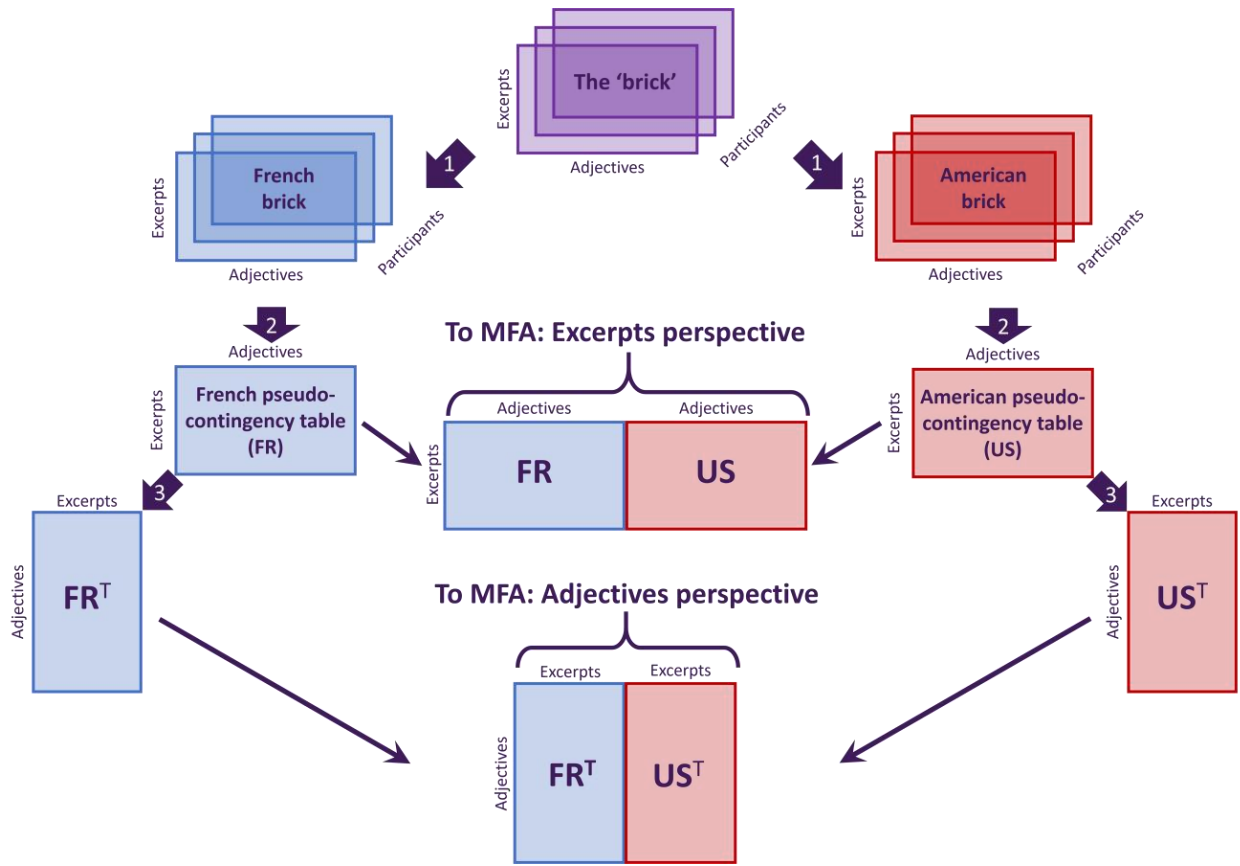

*Note.* 1. Brick separated by nationality. 2. Separate bricks summed across pages. 3. Tables transposed. Thin arrows: tables as blocks concatenated into large matrices and sent to MFA for analysis.

## Results

### *Excerpts*

### *CA & HCA*

The CA performed on the AS pseudo-contingency table revealed two important dimensions, which together accounted for 73% of the total variance (see Figure 6).

Figure 7 displays the factor scores for the excerpts and the adjectives. The interpretation of these plots is similar to the interpretation of the factor plots for Experiment 1. Because this experiment captures participant behavior relative to the descriptions of the excerpts, adjectives

that are near one another can be interpreted as having been used similarly, such as "Incisive" and

"Complex." This plot shows a clear valence-arousal plane, such that the first dimension represents

valence, with adjectives such as "Sad" and "Dark" on the right contrasting with "Dancing" and

"Happy" on the left, and the second dimension represents arousal, such that "Aggressive" is contrasted near the top with "Soft" near the bottom. Similarly, Excerpts 27 and 26 are defined almost entirely by valence, with their projections on Dimension 1 accounting for 84% and 86% of their variance, respectively, and Excerpt 28 could be interpreted as being defined almost entirely by arousal, with its projection on Dimension 2 accounting for 81% of its variance.

*Figure 6.* CA: Scree plot for Adjectives Survey, showing percentage of explained variance per dimension. Horizontal line indicates the average variance extracted per dimension.

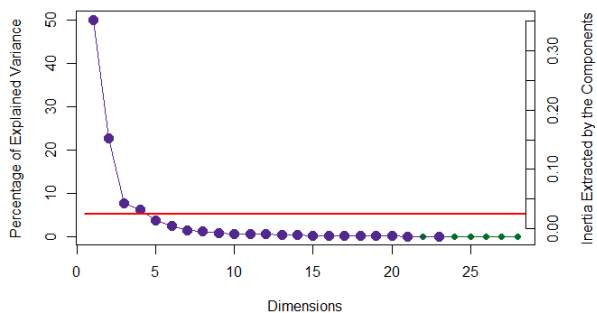

*Figure 7.* CA: Adjective survey, factor plots for Excerpts and Adjectives, each colored according to clusters identified by their respective HCAs. Axis labels indicate dimension, eigenvalue, and explained variance for that dimension.

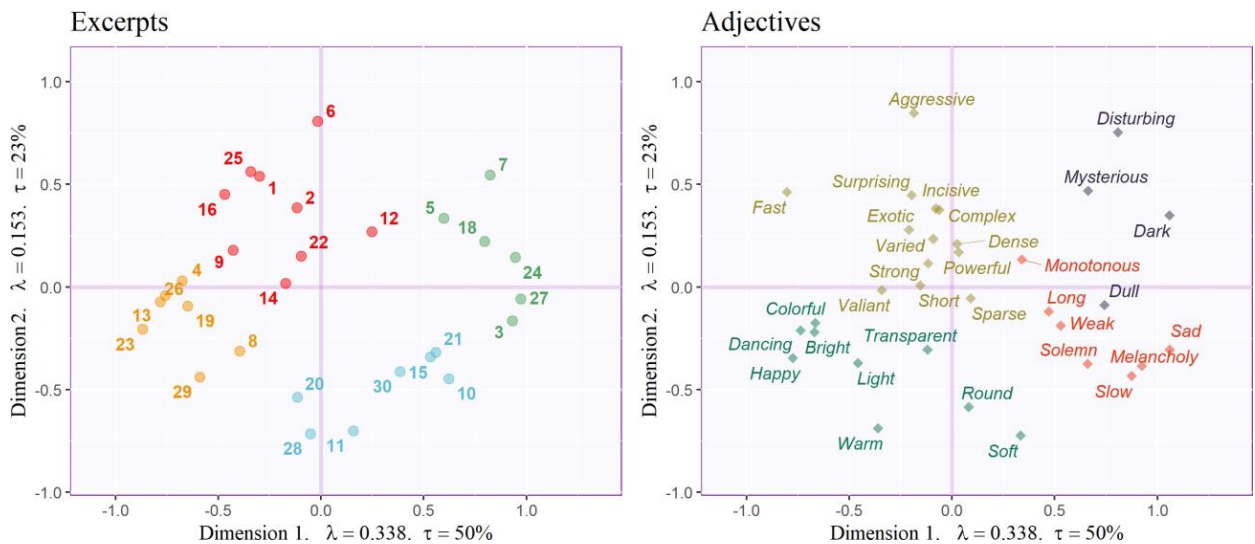

The HCAs computed separately on the factor scores for the rows (excerpts) and columns (adjectives; see supplementary materials for tree diagrams) revealed four clusters for each set, and the excerpts and adjectives are colored according to these clusters for all plots for Experiment 2. The clusters of adjectives and excerpts identified by the HCA are grouped approximately by quadrant in Figure 7, with the top right representing negative valence/high arousal, the top left representing positive valence/high arousal, the bottom left representing positive valence/low arousal, and the bottom right representing negative valence/low arousal. A few adjectives do not conform to this pattern—such as "Monotonous" and "Dull"—because the factor scores for all dimensions were used for the HCA, and these adjectives were likely loading on higher dimensions.

Figure 8 displays the contributions of excerpts and adjectives important for the first two dimensions. For Dimension 1, adjectives that describe negative valence contribute to the positive side, while those that describe positive valence contribute to the negative side. For Dimension 2, adjectives that describe high arousal contribute to the positive side and those that describe low arousal contribute to the negative side. As in Experiment 1, the excerpts that are characterized by these adjectives contribute similarly to their respective dimensions and directions.

*Figure 8.* CA: Adjective survey. Important signed contributions from rows and columns, colored according to clusters identified by their respective HCAs. The *y*-axis represents the value of the contributions.

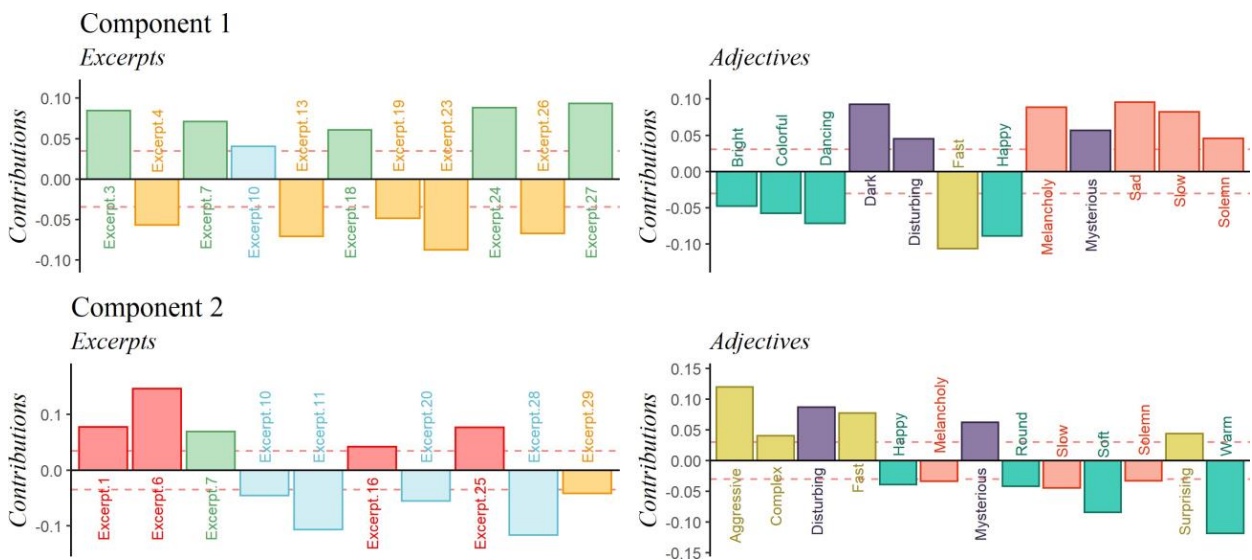

*Participants*

*MDS*

Figure 9 displays the factor scores obtained from the MDS performed on the co-occurrence matrix for the AS, along with group means and bootstrap-derived confidence intervals. The separation between the confidence intervals indicates significant differences between French and American participants ($p < .001$). Group differences between French ($M = 0.04$, $SD = 0.04$) and American ($M = –.03$, $SD = 0.08$) participants are confirmed by the results of a *t*-test on the factor scores on the first dimension, $t(268.89) = 9.63$, $p < .001$. Additional analyses using gender identity and level of music training as factors did not reveal any significant difference.

*Figure 9.* MDS: Distance analysis of the co-occurrence matrix for the adjectives survey, including group means and bootstrap-derived confidence intervals, colored by nationality. Axis labels indicate dimension, eigenvalue, and explained variance for that dimension.

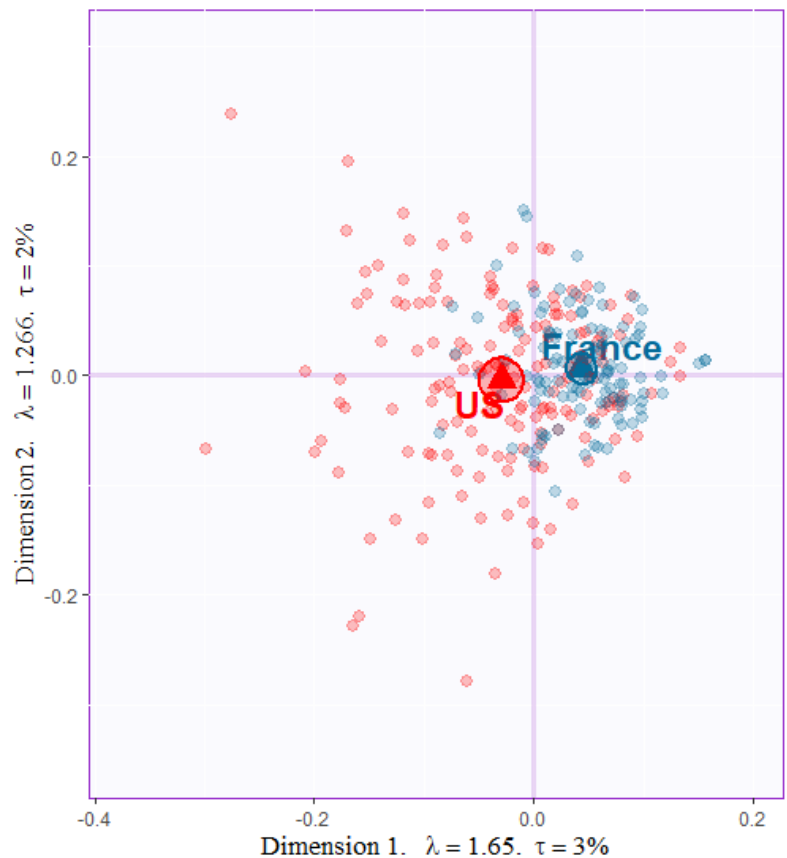

An additional HCA performed post-hoc on the factor scores of the MDS revealed two clusters that somewhat aligned with the *a priori* nationality groupings. One group consisted of 101 French participants (90.2%) and 81 American participants (48.8%), and the other group consisted of 11 French participants (9.8%) and 85 American participants (51.2%). A plot of the MDS results using these results as a grouping variable is included in the supplementary materials.

*MFA*

Figure 10 displays the results of the MFAs as partial factor score plots (see "Multiple Factor Analysis" above) highlighting differences in descriptions from the perspective of the excerpts (left) and the adjectives (right) between French and American participants. The two separate MFAs revealed slightly different factorial dimensions, as shown by the percentage of extracted variance by each axis (denoted τ in the Figures), but the general space for both plots is similar to the space revealed by the CA for Experiment 2 (Figure 7). Thus, we can interpret the space similarly, relative to the valence-arousal plane. However, in this case, we cannot compare elements between maps.

*Figure 10.* Compromise (diamonds) and partial factor scores (small circles) for MFA analyses on the Excerpts and Adjectives, colored according to clusters identified by the respective HCAs. Axis labels include dimension, eigenvalue, and explained variance for that dimension.

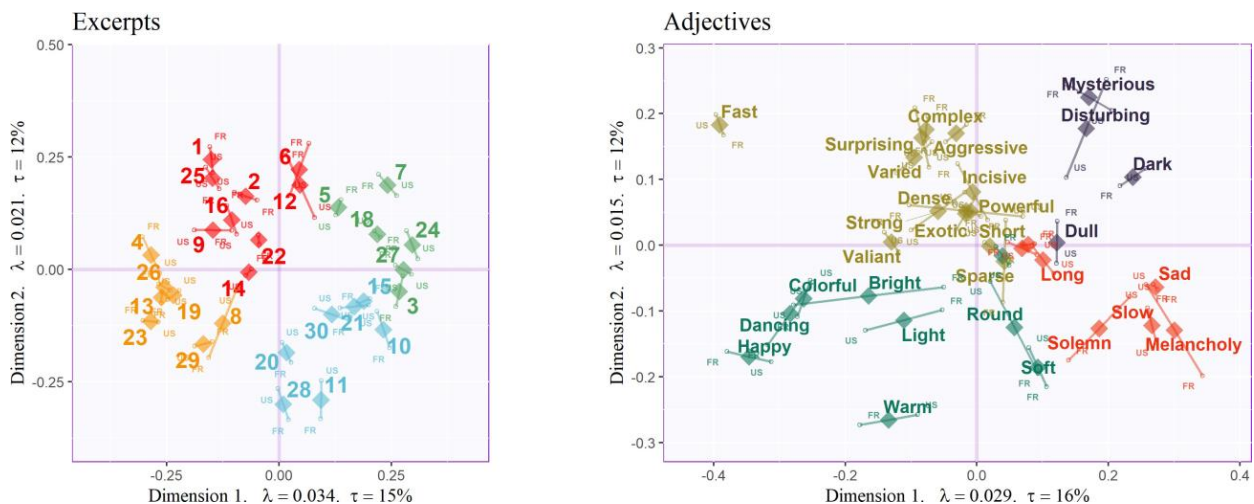

The diamonds represent the compromise between the mental spaces of the French and American participants for each item, and the lines extending from the diamonds to the circles point to the partial factor scores for the items from the perspective of each group (Abdi et al., 2013). Excerpts and adjectives that were rated similarly by each group have short lines extending from them, and those that were rated differently by each group have longer lines. Examples of excerpts that were rated differently are numbers 6, 8, and 12. Adjectives that were used differently include "Disturbing," "Round," "Solemn," and "Bright."

**Experiment 2 Discussion**

As suggested by the MDS on the participants (Figure 9), American and French participants differed in both their mean and their variance. The larger variance of the American participants likely indicates that American participants constitute a more heterogeneous group than French participants. This heterogeneity of the American participants is reflected in their varied responses to the nationality question (with nine different answers), compared to the French participants who all responded "French" (except for one participant who responded with "French – Belgian").

Because participants only rated half of the excerpts, the mean group differences and confidence intervals are meaningful, but the proximity between individual participants cannot be directly interpreted as similarity. A better estimation of between-subject similarity would need to weight the similarity (i.e., the number of common adjectives chosen) between two participants by the number of common excerpts presented.[2]

The adjectives used in Experiment 2 were not selected to engage an affective appraisal,

---

[2] The results of the CA, on the other hand, are not affected by the fact that participants only rated half of the excerpts.

but the first two dimensions of the MFA nevertheless reveal that participants were answering using affective dimensions such as valence and arousal—a result that resonates with previous work investigating conceptual organization (Osgood & Suci, 1955) and music and emotion (Wedin, 1969, 1972).

As a consequence, differences between French and American participants include a large proportion of evaluative adjectives such as "Bright," "Light," "Round," "Solemn," "Melancholy," and "Disturbing." The adjective "Bright" (Brillant) may be the most extreme example of this intercultural difference, as the French partial factor score is close to the origin whereas  the American partial factor score is further away—a difference suggesting that this word has a more positive valence in English than in French. This interpretation is supported by information from the Extended Open Multilingual Wordnet (Bond & Foster, 2013), which shows semantic associations within and across languages. In French, "Brillant" is associated only with physical descriptions of color or light, whereas in English, "Bright" is also associated with happiness or positive qualities like promise. (e.g., "a bright future"). The inverse of "Bright" might be "Round," (Tendre), whose French partial factor score is further from the origin than the American. In this case, the English associations with "Round" include physical descriptions, while the French associations include many more affective references (Bond & Foster, 2013). "Melancholy" (Mélancolique) and "Sad" (Triste) were almost synonymous in English, but not in French.  This difference mirrors early semantic differential and psycholinguistic work that suggests that the usage patterns of adjectives between French and English are different (Osgood et al., 1975).

### Experiment 3:  Combined Surveys

**Justification**

The data obtained in Experiments 1 and 2 capture different aspects of the perception of the excerpts. Experiment 1 asked participants to evaluate musical characteristics, on objective musical dimensions, and Experiment 2 asked participants to evaluate the music subjectively, not using musical characteristics. This method of gathering participant responses on two aspects of the stimuli is similar to that of Balkwill and Thompson (1999), although we differ here in that we use music-theoretical dimensions instead of psychophysical ones. The goal of Experiment 3 was to evaluate which musical characteristics and subjective descriptors are associated with the same excerpts, and therefore with one another.

We acknowledge that we are comparing—in addition to different data—different populations of participants. The participants for Experiment 1 were selected from a population of experts because we used technical terminology that musical novices would not have been familiar with and would probably not know how to use. The participants for Experiment 2 were selected without regard to training because it has been found that musically trained and untrained listeners evaluate music similarly with regard to affect (Bigand & Poulin-Charronnat 2006).

The comparison of these two sets of data is not unlike the procedures used in Music Information Retrieval (MIR) studies, in which participant subjective appraisal is compared to data extracted from the music itself (see Panda et al., 2020 for a review). Although there have been massive strides in the field of MIR in aligning the information extracted by the computer with human perception, there is still a gap between the algorithmic extraction and human perception. It thus can be difficult to identify what information extracted by the computer is perceived by human listeners and vice-versa. However, in comparing two different types of human listener appraisal, we can directly compare these perceivable musical dimensions to the kinds of qualities listeners assign to that music during listening.

**Methods**

Because Experiment 3 used the data tables computed for Experiments 1 and 2 for its analysis—Partial Least Squares Correlation (PLSC)—no additional data collection was necessary. However, because PLSC requires the same sets of observations, and because Experiments 1 and 2 removed different excerpts, we removed from the data the excerpts present in only one table. Specifically, Excerpt 17 was removed from the table used in Experiment 1 and Excerpts 6 and 14 were removed from the table used in Experiment 2. This way, both data tables comprised data from the same 27 Excerpts.

**Results**

The PLSC performed using the pseudo-contingency tables from Experiments 1 and 2 revealed two significant dimensions which, together, accounted for 84.25% of the total variance (shown in Figure 11).

*Figure 11.* PLSC: Scree plot showing explained variance per dimension. Horizontal line represents the average variance extracted per dimension.

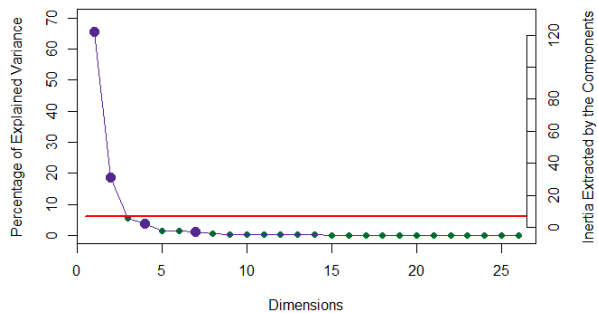

PLSC displays the latent variable from one table against the latent variable from the other table (e.g., LV1 from Table 1 vs. LV1 for Table 2). Figure 12 displays the LVs plot for LVs 1 and 2. In these plots, the excerpts are colored according to the clusters identified by the HCA for Experiment 2, along with tolerance intervals comprising the elements from each cluster. The first

LVs (Figure 12, left) separate the excerpts with positive valence and low arousal (gold) from those with negative valence and high arousal (green). The second LVs (Figure 12, right) separate the groups with positive valence and high arousal (red) from excerpts with negative valence and low arousal (blue).

*Figure 12.* PLSC: Latent variables for Experiment 1 contingency table (horizontal, *x*) plotted against latent variables for Experiment 2 contingency table (vertical, *y*), including tolerance intervals, colored according to the groups revealed by Experiment 2

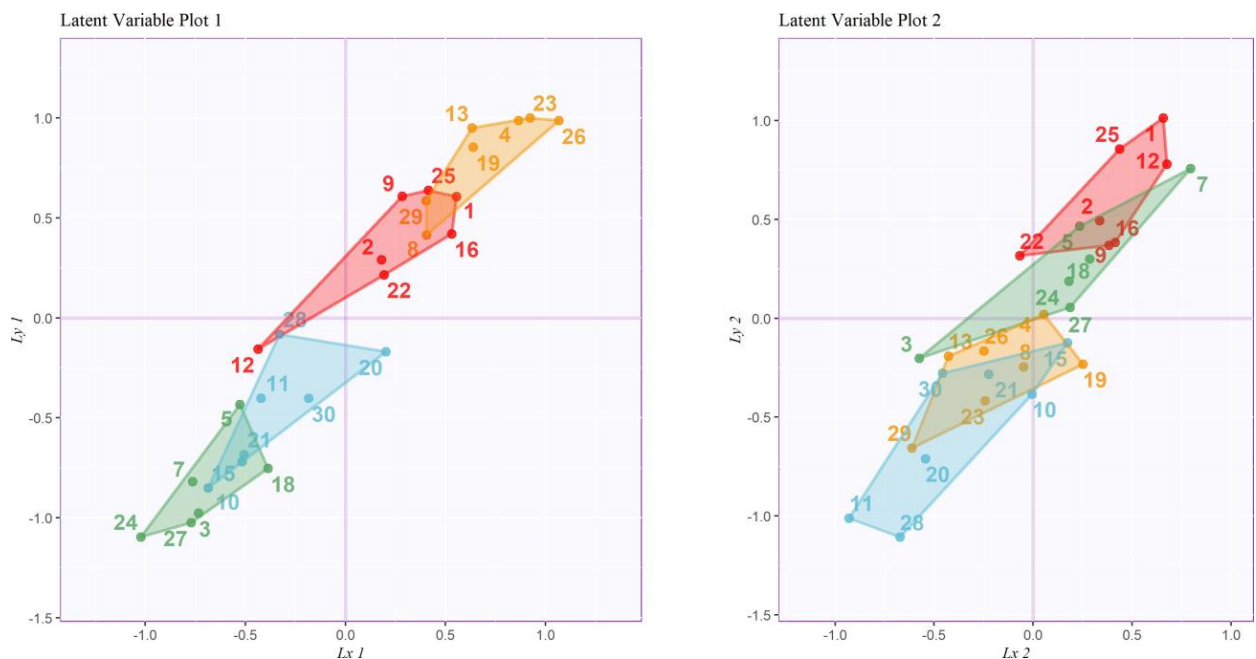

Figure 13 displays the contributions from the variables from each data table that are important for the first and second LVs. For these plots, the important levels of variables from Experiment 1 are displayed in green and the important adjectives from Experiment 2 are in blue. The first LVs from each table feature contributions from levels of variables identified as contributing to an arousal dimension in Experiment 1 and the adjectives identified as contributing to a valence dimension in Experiment 2. The second LVs from each table feature contributions from levels of variables identified as contributing to the genre or complexity dimension from Experiment 1 and adjectives identified as contributing to an arousal dimension in Experiment 2.

*Figure 13*. PLSC: Signed contributions important for the first and second latent variables, colored by survey of origin.

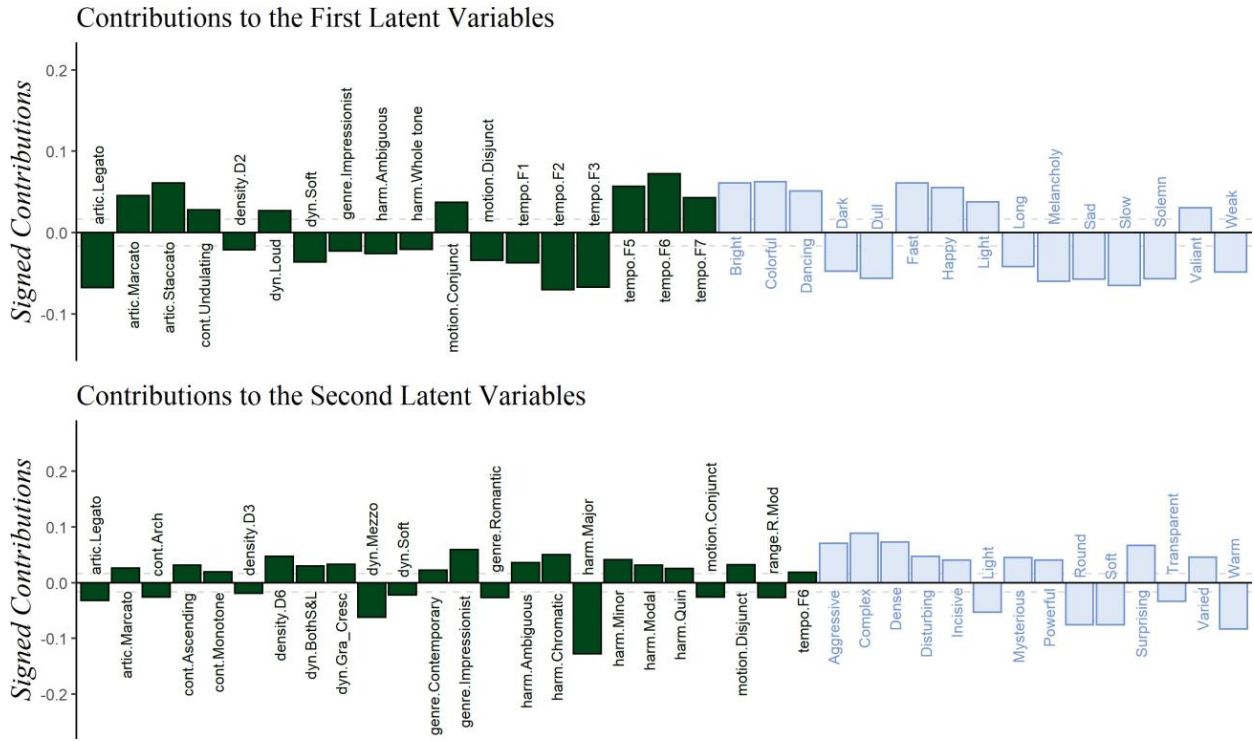

## Experiment 3: Discussion

The goal of this experiment was to identify the common information in the data tables used in Experiments 1 and 2. The first and second latent variables separated the excerpts along dimensions similar to the dimensions extracted by Experiments 1 and 2. Specifically, the first LVs combined the arousal dimension from Experiment 1 with the valence dimension from Experiment 2, and the second LVs combined the complexity or genre dimension from Experiment 1 with the arousal dimension from Experiment 2.

## General Discussion

We collected survey responses to musical stimuli and used multivariate analyses to explore the musical listening spaces created by participants from France and the United States.

The results revealed commonalities and differences between these two national groups. French and American participants agreed on: 1) a clear valence-arousal plane common to participants from both countries when describing the stimuli using adjectives, and 2) a space defined by arousal and complexity when evaluating stimuli using musical qualities. However, French and American participants disagreed on the way in which they used the adjectives when describing the stimuli, a result that suggests either cultural differences in the affective response to the stimuli or, more likely, differences in the use of the adjectives between the two languages.

The results of the MDS analyses across experiments showed group differences between French and American participants when they described excerpts using adjectives (Experiment 2), but these results did not show group differences when experts rated excerpts on specific musical qualities (Experiment 1). This pattern of results suggests that Experiment 1 reveals more about the excerpts themselves rather than about the behavior of participants.

Experiment 3 integrates the results of Experiments 1 and 2, because the first and second latent variables of Experiment 3 essentially integrate the dimensions of Experiments 1 and 2. For example, Excerpt 26—a very distal point in the first LV plot in Figure 12—is an important contributor to the first dimensions of the CAs for both Experiments 1 and 2 (see Figures 4 and 8 for the contributions), but is not an important contributor to the second dimensions of Experiments 1 and 2, and is therefore close to the origin in the second LV plot. By contrast, Excerpt 7—a large contributor to the first and second dimensions of the CAs of both Experiments 1 and 2—is far from the origin in both plots in Figure 12.

The differences in results between Experiments 1 and 2—specifically regarding Excerpts 6 and 14—demonstrate how small differences in experimental paradigm can provide large differences in perspective. In Experiment 1 (by contrast with Experiment 2), the experts rating the excerpts on specific musical qualities isolated two Excerpts: 6—a minimalist, ostinato based excerpt—and 14—a jazzy excerpt, each the only representative of their style. There are a few possible reasons for this pattern of results, including differences 1) in participant characteristics— experts in Experiment 1 versus non-experts in Experiment 2—and 2) in the way the way the questions in each survey assessed the excerpts—with specific musical qualities in Experiment 1 and subjective evaluations in Experiment 2. Of these two interpretations, the second is more likely, because the (few) participants in Experiment 2 with significant musical training did not differ in their descriptions of the excerpts from the untrained participants.

The different experimental paradigms in the present study all provide useful perspectives. For example, the paradigm from Experiment 1—which separates stimuli along concrete musical dimensions—effectively reveals stimulus differences, whereas the paradigm from Experiment 2 reveals stimulus affective similarity. In addition, the combination of these two paradigms (as in Experiment 3) probes the "why" of the stimulus affective impact.

**Why these methods?**

Whereas many readers may already be familiar with such methods as MDS or HCA (which are commonly used in many domains), one goal of the present work was to present less familiar options—such as CA, MFA, and PLSC—for consideration. Table 3 displays these methods along with similar methods and possible applications. Because each analysis offers a different perspective or is best suited to handle a specific type or shape of data, familiarity with a range of analyses is useful both when approaching existing questions and exploring new directions.

As stated above, CA is similar to PCA, but can be performed using qualitative data, which makes it a valuable addition to any qualitative analysis. Also, if a research question would benefit

Table 3

*Methods and their uses.*

| Method | Some Similar methods | Kind of data | Useful for |
|---|---|---|---|
| Correspondence Analysis (CA) | Latent Semantic Analysis (LSA) Discriminant Correspondence Analysis (DICA) Multiple Correspondence Analysis (MCA) Canonical Correspondence Analysis | Qualitative, as a contingency table or pseudo-contingency table | Visualizing sets of observations and variables in the same space. A number of extensions of CA, including Discriminant Correspondence Analysis (DiCA) can provide additional inferences. |
| Hierarchical Cluster Analysis (HCA) | Additive tree clustering MDS | Sorting data, distance matrices, data that represent classification or ordination in some way | Identifying clusters or groups within the data that may not be identified a priori. If the data are a contingency table, this can be used to identify clusters of variables or observations. If the data are a distance matrix or similar, this can identify clusters of items on which distance is being measured. |
| Metric Multidimensional Scaling (MDS) | PCA DISTATIS Non Metric Multidimensional Scaling (NMMDS) HCA | Distance matrices, Confusion matrices, matrices of correlations, sorting data | Evaluating similarity or dissimilarity between observations, variables, participants, or groups. Visualizes distance on a plane. |
| Multiple Factor Analysis (MFA) | PCA DISTATIS STATIS | Multiple data tables (not limited to two), each with observations obtained on the same set of variables or vice-versa. | Visualizing how groups of observations have different perspectives on the variables. If the data are a contingency or pseudo-contingency table, the tables can be transposed to visualize the observations. |
| Partial Least Squares Correlation (PLSC) | PLSCA PLSR Canonical Correlation Analysis | Two data tables with the same observations (rows), that may have different variables. Could also be the same set of variables taken at a different time, for example. | Used in brain imaging to evaluate what brain regions (as voxels, table one) are active during cognitive tasks (as performance scores, table two). Generalizable to any two sets or groups of variables gathered on a set of observations, to see what information is shared. |

from visualizing variables *and* observations in the same space, CA is the method of choice. Biplots—that is, both plots on a single set of axes—were not used for the plots above for clarity, given the space and font size constraints.

MFA is, conceptually and practically, an exploratory method. Its strength lays in the partial factor scores revealing how groups of participants, products, or stimuli have different perspectives on variables or observations. These groups might be defined *a priori* or could be determined *a posteriori* by an HCA or similar method. Although we only used two groups in the present study, MFA is not limited to two groups or data sets—the number of groups is only limited by interpretability, as long as the variables measured for each group of observations are the same.

PLSC is commonly used in fMRI analysis to find which brain regions are active during behavioral tasks. However, this is only one possible use of this technique—it was initially developed for econometrics and chemometrics (Wold, 1982). As we show above, it can be used to identify what information is shared between two datasets, even when the shared information comprises some previously unidentified variables, or in a situation that is "data-rich and theory-skeletal" (Wold, 1982). We urge caution, however, against applying this method indiscriminately, because the data common to the two tables may be spurious, as described by Bennett et al. (2011).

Readers who are curious about the qualitative differences between MFA and PLSC are encouraged to review figures 10 and 12. Figure 10 shows how MFA is better suited for showing group perspectives on the *existing* variables via the partial factor scores. Figure 12 (PLSC) shows how the latent variables identify shared information between the two datasets that may not be apparent in the original data—thus PLSC is ideal when the research question involves identifying underlying structures or tertiary variables in the data. However, not shown in these two figures is the fact that MFA is usable with three or more data sets, while PLSC is limited to two.

MDS and HCA are similar analyses because they evaluate similarity between items. However, the outputs of these methods offer different perspectives on the data. For example, MDS is best suited to provide an intuitive visualization of similarity as proximity. The distance visualization provided by HCA is not as intuitive as that of MDS, but it is better for identifying clusters and can help researchers make choices about those clusters when the configuration between points in an MDS plot is unclear.

**Limitations & future directions**

One major difficulty in online data collection is attrition. As we mentioned in the introduction, in the lab, precise control over conditions allows for a small number of participants, and the likelihood of usable data from every participant is much higher. In online data collection, because there is no control over whether the participant finishes, follows the experimental protocol, or even answers in good faith, much of the data may be incomplete. In Experiment 1, for example, only 32% of responses were usable. Many of these responses appeared to be participants who followed the link to the survey and accepted the consent form but did not start the survey. It is unclear whether any of these responses are from individuals who opened the survey multiple times and only completed it once or simply read through the form and then decided not to participate. The tradeoff, of course, is that it is easier to collect a larger volume of data, especially from participants who might not otherwise be accessible.

Because the participants in Experiment 1 self-identified as only French or American, excluding participants who did not identify as American or an American-other nationality compound in Experiment 2 was necessary to control the comparison between participants in Experiments 1 and 2. A separate MDS analysis was performed on the data including the excluded participants as a third group. This analysis revealed similar differences between the third group and the French participants as between the American and French participants, however, no significant differences were revealed between the US participants who identified as American and those who did not. This highlights the fact that nationality is an imperfect surrogate for culture or language, especially in a diverse environment. It also indicates how recruitment and data cleaning procedures need to be robust to collect enough data that there is enough data to analyze after attrition.

Although we evaluated scores and ratings of participants from different countries, we did not explicitly address multiculturality, because France and the United States are both Western countries that share the same Western musical culture. To address this multicultural question, an experiment would need to include music and/or participants from multiple and contrasted musical cultures. However, specific musical qualities, such as harmony, may not apply or translate well to other musical cultures, because the concepts of melodic and harmonic material are not the same across all musical cultures (Cohn et al., 2001; Raman & Dowling, 2017). We also suggest that data collected in this way have a much greater hypothetical reach, but the data collected for these experiments represent a convenience sample, and many of the participants were students. However, this limitation could be easily remedied in future studies.

**Conclusions**

On-line data collection and multivariate analysis are not simply a palliative to be used in a time of pandemic. In fact, this paradigm not only enriches the psychologist's methodological tool-box, but it also may be one of the best ways of reaching a more representative population than first year undergraduate students in psychology.

**References**

Abdi, H. (2007). Metric Multidimensional Scaling. In N.J. Salkind (Ed.): *Encyclopedia of Measurement and Statistics*. Sage.

Abdi, H. (2020). PTCA4CATA: Partial triadic analysis for check all that apply (CATA) data. http://github.com/HerveAbdi/PTCA4CATA

Abdi, H., & Béra, M. (2018). Correspondence analysis. In R. Alhajj & J. Rokne (Eds.), *Encyclopedia of social networks and mining* (2nd ed., pp. 275–284). Springer Verlag. https://doi.org/10.1007/978-3-642-04898-2_195

Abdi, H., & Williams, L. J. (2010). Correspondence analysis. In N. Salkind (Ed.), *Encyclopedia of research design*. Sage.

Abdi, H., & Williams, L. J. (2013). Partial least squares methods: Partial least squares correlation and partial least square regression. In B. Reisfeld & A. N. Mayeno (Eds.), *Methods in molecular biology: Computational toxicology volume II* (Vol. 930, pp. 549–579). Springer Science+Business Media, LLC. https://doi.org/10.1007/978-1-62703-059-5

Abdi, H., Williams, L. J., & Valentin, D. (2013). Multiple factor analysis: Principal component analysis for multitable and multiblock data sets. *Wiley Interdisciplinary Reviews: Computational Statistics*, *5*, 149–179. https://doi.org/10.1002/wics.1246

Ares, G., Deliza, R., Barreiro, C., Giménez, A., & Gámbaro, A. (2010). Comparison of two sensory profiling techniques based on consumer perception. *Food Quality and Preference*, *21*(4), 417–426. https://doi.org/10.1016/j.foodqual.2009.10.006

Auguie, B. (2017). gridExtra: Miscellaneous functions for "grid" graphics. https://CRAN.R-project.org/package=gridExtra

Aust, F., & Barth, M. (2020). papaja: Create APA manuscripts with R Markdown.

https://github.com/crsh/papaja

Balkwill, L. L., & Thompson, W. F. (1999). A cross-cultural investigation of the perception of emotion in music: Psychophysical and cultural cues. *Music Perception: An Interdisciplinary Journal*, *17* (1), 43–64. https://doi.org/10.2307/40285811

Balkwill, L. L., Thompson, W. F., & Matsunaga, R. (2004). Recognition of emotion in Japanese, Western, and Hindustani music by Japanese listeners. *Japanese Psychological Research*, *46*(4), 337–349. https://doi.org/10.1111/j.1468-5584.2004.00265.x

Bartlett, J. C., & Dowling, W. J. (1980). Recognition of transposed melodies: A key-distance effect in developmental perspective. *Journal of Experimental Psychology: Human Perception and Performance*, *6*(3), 501–515. https://doi.org/10.1037/0096-1523.6.3.501

Battcock, A., & Schutz, M. (2019). Acoustically expressing affect. *Music Perception*, *37* (1), 66–91. https://doi.org/10.1525/MP.2019.37.1.66

Beaton, D., Fatt, C. R. C., & Abdi, H. (2014). An ExPosition of multivariate analysis with the singular value decomposition in R. *Computational Statistics & Data Analysis*, 72(0), 176–189. https://doi.org/10.1016/j.csda.2013.11.006

Bennett, C. M., Baird, A. A., Miller, M. B., & Wolford, G. L. (2011). Neural correlates of interspecies perspective taking in the post-mortem Atlantic salmon: An argument for proper multiple comparisons correction. *Journal of Serendipitous and Unexpected Results*, *1*(1), 1-5.

Benzécri, J.-P. (1973). *L'analyse des données.* Dunod.

Berry, K. J., Kvamme, K. L., Johnston, J. E., & Mielke, P. W. (2021). *Permutation Statistical Methods with R*. Springer. 10.1007/978-3-319-98926-6_2

Beygelzimer, A., Kakadet, S., Langford, J., Arya, S., Mount, D., & Li, S. (2019). FNN: Fast

nearest neighbor search algorithms and applications. https://CRAN.R-project.org/package=FNN

Bigand, E., & Poulin-Charronnat, B. (2006). Are we "experienced listeners"? A review of the musical capacities that do not depend on formal musical training. *Cognition*, *100*(1), 100–130. https://doi.org/10.1016/j.cognition.2005.11.007

Bigand, E., Vieillard, S., Madurell, F., Marozeau, J., & Dacquet, A. (2005). Multidimensional scaling of emotional responses to music: The effect of musical expertise and of the duration of the excerpts. *Cognition and Emotion*, *19*(8), 1113–1139. https://doi.org/10.1080/02699930500204250

Borg, I., & Groenen, P. J. F. (2005). *Modern Multidimensional Scaling* (2nd ed., Vol. 36). Springer.

Bond, F., & Foster, R. (2013, August 4-9). *Linking and extending an open multilingual wordnet*. Proceedings of the 51st Annual Meeting of the Association for Computational Linguistics, Sofia. aclweb.org/anthology/P/P13/P13-1133.pdf

Bruner II, G. C. (1990). Music, mood, and marketing. *Journal of Marketing*, (October), 94–104.

Cohn, R., Hyer, B., Dahlhaus, C., Anderson, J., & Wilson, C. (2001). *Harmony*. Oxford University Press.

Coombs, C. H., Milholland, J. E., & Womer, F. B. (1956). The assessment of partial knowledge. *Educational and Psychological Measurement*, *16*(1), 13–37. https://doi.org/10.1177/001316445601600102

Coombs, C.H. (1964). *A theory of data.* Wiley.

Cowen, A. S., Fang, X., Sauter, D., & Keltner, D. (2020). What music makes us feel: At least 13 dimensions organize subjective experiences associated with music across different cultures. *Proceedings of the National Academy of Sciences of the United States of America*, *117* (4), 1924–1934. https://doi.org/10.1073/pnas.1910704117

Daróczi, G., & Tsegelskyi, R. (2018). Pander: An R 'pandoc' writer. https://CRAN.R-project.org/package=pander

Darrow, A. A., Haack, P., & Kuribayashi, F. (1987). Descriptors and preferences for eastern and western musics by Japanese and American nonmusic majors. *Journal of Research in Music Education*, *35*(4), 237–248. https://doi.org/10.2307/3345076

Derek Beaton, H. A. &. (2020). data4PCCAR: Some data sets and r-functions for PCA and CA to accompany Abdi & Beaton (2019) principal component and correspondence analysis with r. http://github.com/HerveAbdi/data4PCCAR

Dowle, M., & Srinivasan, A. (2020). Data.table: Extension of 'data.frame'. https://CRAN.R-project.org/package=data.table

Dowling, W. J. (1978). Scale and contour: Two components of a theory of memory for melodies. *Psychological Review*, *85*(4), 341–354. https://doi.org/10.1037/0033-295X.85.4.341

Escofier, B., & Pagès, J. (1994). Multiple factor analysis (AFMULT package). *Computational Statistics and Data Analysis*, *18*(1), 121–140. https://doi.org/10.1016/0167-9473(94)90135-X

Escofier-Cordier, B. (1965). *L'analyse des correspondances* [Unpublished doctoral dissertation]. Université de Rennes.

Fatt, C. R. C., Beaton, D., & Abdi., H. (2013). MExPosition: Multi-table ExPosition. https://CRAN.R-project.org/package=MExPosition

Fritz, T., Jentschke, S., Gosselin, N., Sammler, D., Peretz, I., Turner, R., Friederici, A. D., & Koelsch, S. (2009). Universal recognition of three basic emotions in music. *Current*

*Biology*, *19*(7), 573–576. https://doi.org/10.1016/j.cub.2009.02.058

Gagolewski, M. (2020). R package stringi: Character string processing facilities. http://www.gagolewski.com/software/stringi/

Gohel, D. (2021). Flextable: Functions for tabular reporting. https://CRAN.R-project.org/package=flextable

Gower, J. C. (1966). Some distance properties of latent root and vector methods used in multivariate analysis. *Biometrika*, *53*(3/4), 325. https://doi.org/10.2307/2333639

Gray, P. H., & Wheeler, G. E. (1967). The semantic differential as an instrument to examine the recent folksong movement. *Journal of Social Psychology*, *72*(2), 241–247. https://doi.org/10.1080/00224545.1967.9922321

Greenacre, M. J. (1984). *Theory and applications of correspondence analysis*. Academic Press.

Gregory, A. H., & Varney, N. (1996). Cross-cultural comparisons in the affective response to music. *Psychology of Music*, *24*(1), 47–52. https://doi.org/10.1177/0305735696241005

Henry, L., & Wickham, H. (2020). Purrr: Functional programming tools. https://CRAN.R-project.org/package=purrr

Hester, J. (2020). Glue: Interpreted string literals. https://CRAN.R-project.org/package=glue

Hesterberg, T. (2011). Bootstrap. *Wiley Interdisciplinary Reviews: Computational Statistics*, *3*(6), 497–526. https://doi.org/10.1002/wics.182

Hout, M. C., Papesh, M. H., & Goldinger, S. D. (2013). Multidimensional scaling. *WIREs Cognitive Science*, *4*(1), 93-103. 10.1002/wcs.1203

Juslin, P. N., & Sloboda, J. A. (Eds.). (2010). *Handbook of music and emotion: Theory, research, applications.* Oxford University Press.

Juslin, P. N., & Västfjäll, D. (2008). All emotions are not created equal: Reaching beyond the traditional disputes. *Behavioral and Brain Sciences*, *31*, 559–621.

https://doi.org/doi:10.1017/S0140525X08005554%20Patrik

Katz, D., & Braly, K. (1933). Racial stereotypes of one hundred college students. *Journal of Abnormal and Social Psychology*, *28*(3), 280–290. https://doi.org/10.1037/h0074049

Kopacz, M. (2005). Personality and music preferences: The influence of personality traits on preferences regarding musical elements. *Journal of Music Therapy*, *42*(3), 216–239. https://doi.org/10.1093/jmt/42.3.216

Krishnan, A., Williams, L. J., McIntosh, A. R., & Abdi, H. (2011). Partial least squares (PLS) methods for neuroimaging: A tutorial and review. *NeuroImage*, *56*(2), 455–475. https://doi.org/10.1016/j.neuroimage.2010.07.034

Kruskal, J. B. & Wish. M. (1978). *Multidimensional Scaling*. Sage Publications.

Ladinig, O., & Schellenberg, E. G. (2012). Liking unfamiliar music: Effects of felt emotion and individual differences. *Psychology of Aesthetics, Creativity, and the Arts*, *6*(2), 146–154. https://doi.org/10.1037/a0024671

Madsen, C. K. (1997). Emotional response to music as measured by the two-dimensional CRDI. *Journal of Music Therapy*, *34*(3), 187–199. https://doi.org/10.1093/jmt/34.3.187

Meyners, M., & Castura, J. (2014). Check-all-that-apply questions. In *Novel techniques in sensory characterization and consumer profiling* (pp. 271–306). CRC Press/Taylor & Francis. https://doi.org/10.1201/b16853-12

Müller, K., & Wickham, H. (2020). Tibble: Simple data frames. https://CRAN.R-project.org/package=tibble

Osgood, C. E., May, W. H., & Miron, M. S. (1975). *Cross-Cultural Universals of Affective Meaning*. University of Illinois Press.

Osgood, C. E., & Suci, G. J. (1955). Factor analysis of meaning. *Journal of Experimental Psychology*, *50*(5), 325–338. https://doi.org/10.1037/h0043965

Panda, R., Malhiero, R., & Paiva, R. P. (2020) Novel audio features for music emotion recognition. *IEEE Transactions on Affective Computing*, *11*(3), 614-626. 10.1109/TAFFC.2018.2820691

Pielou, E. C. (1984). *The Interpretation of Ecological Data: A Primer on Classification and Ordination*. Wiley.

Plate, T., & Heiberger, R. (2016). Abind: Combine multidimensional arrays. https://CRAN.R-project.org/package=abind

R Core Team. (2020). R: A language and environment for statistical computing. R Foundation for Statistical Computing. https://www.R-project.org/

Ram, K., & Wickham, H. (2018). Wesanderson: A wes anderson palette generator. https://CRAN.R-project.org/package=wesanderson

Raman, R., & Dowling, W. J. (2016). Real-time probing of modulations in south indian classical (carnãtic) music by Indian and Western Musicians. *Music Perception*, *33*(3), 367–393. https://doi.org/10.1525/MP.2016.33.03.367

Raman, R., & Dowling, W. J. (2017). Perception of modulations in south indian classical (carnãtic) music by student and teacher musicians: A cross-cultural study. *Music Perception*, *34*(4), 424–437.

Ren, K. (2016). Rlist: A toolbox for non-tabular data manipulation. https://CRAN.R-project.org/package=rlist

Rodà, A., Canazza, S., & De Poli, G. (2014). Clustering affective qualities of classical music: Beyond the valence-arousal plane. *IEEE Transactions on Affective Computing*, *5*(4), 364–

376. https://doi.org/10.1109/TAFFC.2014.2343222

Shepard, R. (1962). The analysis of proximities Multidimensional scaling with an unknown distance function. I. *Psychometrika*, *27*, 125-140. 10.1007/BF02289630

Shepard, R. (1980). Multidimensional scaling, tree-fitting, and clustering. *Science*, *210*(4468), 390-398.

Stanley, D. (2021). apaTables: Create american psychological association (APA) style tables. https://CRAN.R-project.org/package=apaTables

Thompson, B., Roberts, S. G., & Lupyan, G. (2020). Cultural influences on word meanings revealed through large-scale semantic alignment. *Nature Human Behaviour*, *4*(10), 1029–1038. https://doi.org/10.1038/s41562-020-0924-8

Thompson, W. F. (1994). Sensitivity to combinations of musical parameters: Pitch with duration, and pitch pattern with durational pattern. *Perception & Psychophysics*, *56*(3), 363–374. https://doi.org/10.3758/BF03209770

Torgerson, W. S. (1958). *Theory and methods of scaling*. Wiley.

Tucker, L. R. (1958). An inter-battery method of factor analysis. *Psychometrika*, *23*(2), 111–136. https://doi.org/10.1007/BF02289009

Valentin, D., Chollet, S., Lelièvre, M., & Abdi, H. (2012). Quick and dirty but still pretty good: A review of new descriptive methods in food science. *International Journal of Food Science & Technology*, 1–16. https://doi.org/10.1111/j.1365-2621.2012.03022.x

Wallmark, Z. (2019). A corpus analysis of timbre semantics in orchestration treatises. *Psychology of Music*, *47* (4), 585–605. https://doi.org/10.1177/0305735618768102

Wedin, L. (1969). Dimension analysis of emotional expression in music. *Swedish Journal of Musicology*, *51*, 119–140.

Wedin, L. (1972). Evaluation of a three-dimensional model of emotional expression in music. *The Psychological Laboratories*, *54*(349), 115–131.

Wei, T., & Simko, V. (2017). R package "corrplot": Visualization of a correlation matrix. https://github.com/taiyun/corrplot

Wickham, H. (2019). Stringr: Simple, consistent wrappers for common string operations. https://CRAN.R-project.org/package=stringr

Wickham, H. (2020). Forcats: Tools for working with categorical variables (factors). https://CRAN.R-project.org/package=forcats

Wickham, H., Averick, M., Bryan, J., Chang, W., McGowan, L. D., François, R., Grolemund, G., Hayes, A., Henry, L., Hester, J., Kuhn, M., Pedersen, T. L., Miller, E., Bache, S. M., Müller, K., Ooms, J., Robinson, D., Seidel, D. P., Spinu, V., … Yutani, H. (2019). Welcome to the tidyverse. *Journal of Open Source Software*, *4*(43), 1686. https://doi.org/10.21105/joss.01686

Wickham, H., & Bryan, J. (2019). Readxl: Read excel files. https://CRAN.R-project.org/package=readxl

Wickham, H., François, R., Henry, L., & Müller, K. (2020). Dplyr: A grammar of data manipulation. https://CRAN.R-project.org/package=dplyr

Wickham, H., & Henry, L. (2020). Tidyr: Tidy messy data. https://CRAN.R-project.org/package=tidyr

Wickham, H., Hester, J., & Francois, R. (2018). Readr: Read rectangular text data. https://CRAN.R-project.org/package=readr

Wold, H. (1982). Soft modelling, the basic design and some extensions. In: Wold., H., Jöreskog, K.-G. (Eds.), *Systems Under Indirect Observation: Causality-Structure-Prediction. Part II* (pp. 1-54). North-Holland Publishing Company.

Zacharakis, A., Pastiadis, K., & Reiss, J. D. (2014). An interlanguage study of musical timbre semantic dimensions and their acoustic correlates. *Music Perception: An Interdisciplinary Journal*, *31*(4), 339–358. https://doi.org/10.1525/MP.2014.31.4.339

Zacharakis, A., Pastiadis, K., & Reiss, J. D. (2015). An interlanguage unification of musical timbre: Bridging semantic, perceptual, and acoustic dimensions. *Music Perception: An Interdisciplinary Journal*, *32*(4), 394–412. https://doi.org/10.1525/MP.2015.32.4.394

Zampini, M., & Spence, C. (2004). The role of auditory cues in modulating the perceived crispness and staleness of potato chips. *Journal of Sensory Studies*, *19*(5), 347–363. https://doi.org/10.1111/j.1745-459x.2004.080403.x

Zhu, H. (2020). kableExtra: Construct complex table with 'kable' and pipe syntax. https://CRAN.R-project.org/package=kableExtra